\documentclass[twocolumn,floatfix,aps]{revtex4-1}

\usepackage{graphicx}
\usepackage{amsmath}
\usepackage{amssymb}
\usepackage{bm}
\usepackage{color}
\usepackage{comment}
\usepackage[FIGTOPCAP]{subfigure}
\usepackage[]{footmisc}
\usepackage{lineno}
\usepackage{dsfont}
\usepackage{soul,xcolor}
\usepackage[pdfborder={0 0 0}]{hyperref}

\newcommand{\ts}{\mathrm{s}}
\newcommand{\tss}{\mathrm{s \bar s}}
\newcommand{\tbb}{\mathrm{b \bar b}}
\newcommand{\tb}{\mathrm{b}}
\newcommand{\tsb}{\mathrm{sb}}
\newcommand{\ls}{l^*}

\begin{document}

\title{Large contribution of Fermi arcs to the conductivity of topological metals}

\author{M.\ Breitkreiz}
\email{breitkr@physik.fu-berlin.de}
\author{P.\ W.\ Brouwer}
\affiliation{Dahlem Center for Complex Quantum Systems and Fachbereich Physik, Freie Universit\" at Berlin, 14195 Berlin, Germany}

\date{March 2019}

\begin{abstract}
Surface-state contributions to the dc conductivity of most homogeneous metals exposed to uniform electric fields
are usually as small as the system size is large compared to the lattice constant. In this work, we show that
surface states of topological metals can contribute with the same order of magnitude as the bulk even in large systems. 
This effect is intimately related to the intrinsic anomalous Hall effect, in which an applied voltage induces 
chiral surface-state currents proportional to the system size. Unlike the anomalous Hall effect, the large contribution of surface 
states to the dc conductivity
is also present in time-reversal invariant Weyl semimetals, 
where the surface states come in counter-propagating time-reversed pairs. 
While the Hall voltage vanishes in the presence of time-reversal symmetry, the twinned chiral surface currents develop similarly as in the 
time-reversal broken case. For this effect to occur, the relaxation length associated with scattering between time-reversed partner states needs to be larger than the separation of contributing surfaces, which results in a characteristic size dependence of the resistivity
 and a highly inhomogeneous current-density profile across the sample.
\end{abstract}
\maketitle

\textit{Introduction.}--- Weyl and Dirac semimetals have attracted enormous attention recently, 
especially after their 
experimental discovery four years ago \cite{Xu2015a, Xu2015b, Lv2015, Borisenko2014, Neupane2014, Liu2014a, Xiong2015a}. The fascination for these materials 
 derives in large part from novel responses to electromagnetic fields 
 \cite{Armitage2017, Yan2017}. 
Some of the most striking phenomena are associated with magnetic 
Weyl metals, which in their minimal model consist of
two Weyl nodes close to the Fermi level.
For a strictly linear dispersion around the Weyl points and with the nodes being separated 
by a distance $k_0$ in the $k_z$ direction, the $xy$ part of the conductivity tensor for 
this system is \cite{Burkov2011}
\begin{equation}
\sigma_0 = e^2 \begin{pmatrix} n_\tb D
&\frac{k_0}{2\pi h} \\ -\frac{k_0}{2\pi h} & n_\tb D 
\end{pmatrix},
\label{cond0}
\end{equation}
where $n_\tb$ is the density of bulk states at the Fermi level, $D$ is the bulk diffusion constant, 
and $h$ is Plank's constant.
While the diagonal part follows the standard Drude relation, the off-diagonal part
stems from Berry-curvature singularities at the two Weyl points, which give wave packets an anomalous contribution to their
velocity perpendicular to the applied electric field and the direction of intrinsic magnetization (the $z$ direction) \cite{Xiao2010}. 
This is the anomalous Hall effect --- a transverse conductivity in the absence of an external magnetic field. 
The ratio $\sigma_0^{xy}/\sigma_0^{xx}$ becomes large when Weyl fermions are the only mobile charge carriers
 and the Weyl nodes are close to the Fermi level, in which case the density of bulk states goes to zero.
Experimentally, strong anomalous Hall effects with $\sigma_0^{xy}/\sigma_0^{xx}\sim 0.2$ have been  
observed in GdPtBi \cite{Suzuki2016}  and Co$_3$Sn$_2$S$_2$ \cite{Liu2017c}, 
demonstrating topological Weyl physics in these materials via the associated  
anomalous transport properties.

In this regard it is unfortunate that most existing Weyl metals are time-reversal (TR) 
invariant \cite{Jia2016}. 
While
Berry-curvature singularities and anomalous velocities are equally present in the presence of
TR symmetry, these materials lack the striking  signature of the anomalous Hall voltage. 
 One way to see this is to consider a TR-invariant Weyl metal as the sum of a TR-broken subsystem and its TR-conjugate. Denoting the conductivity tensors of the two subsystems by $\sigma_0$ and $\sigma_0^{\rm T}$, the
total conductivity is then given by the sum $\sigma_0+\sigma_0^{\rm T} = 
2 \sigma_0^{xx}\,\mathds{1}$,  which misses the anomalous off-diagonal part.

Here we will show that the total conductivity is no longer given by $\sigma_0+\sigma_0^{\rm T}$ 
if the relaxation length $\bar l$ associated with scattering between the time-reversed subsystems is
comparable to or larger than the transverse system size $W$. In the limit $\bar l \gg W$ of fully 
decoupled TR subsystems we find that the conductivity is $2 (\rho_{0}^{xx})^{-1}$, 
where $\rho_{0}^{xx}$ is the longitudinal 
resistivity of a single TR subsystem, which is the same for the two TR subsystems. Since $\rho_{0}^{xx} = 
\sigma_{0}^{xx}/[(\sigma_{0}^{xx})^{2} + (\sigma_{0}^{xy})^{2}]$, the anomalous Hall 
conductivity $\sigma_0^{xy}$ of each subsystem does not cancel from the expression for 
the conductivity, but reappears in the diagonal 
part. The additional contribution to the longitudinal conductivity comes, 
as we will show, from topological surface states, which contribute to the conductivity
with the same order of magnitude as bulk states if $\bar l \gtrsim W$. 
Before deriving this result and discussing how the conductivity depends on 
 relevant scattering lengths, we find it useful to first reconsider the anomalous Hall 
effect for a finite-size system.

\begin{figure}[b]
\includegraphics[scale=0.5]{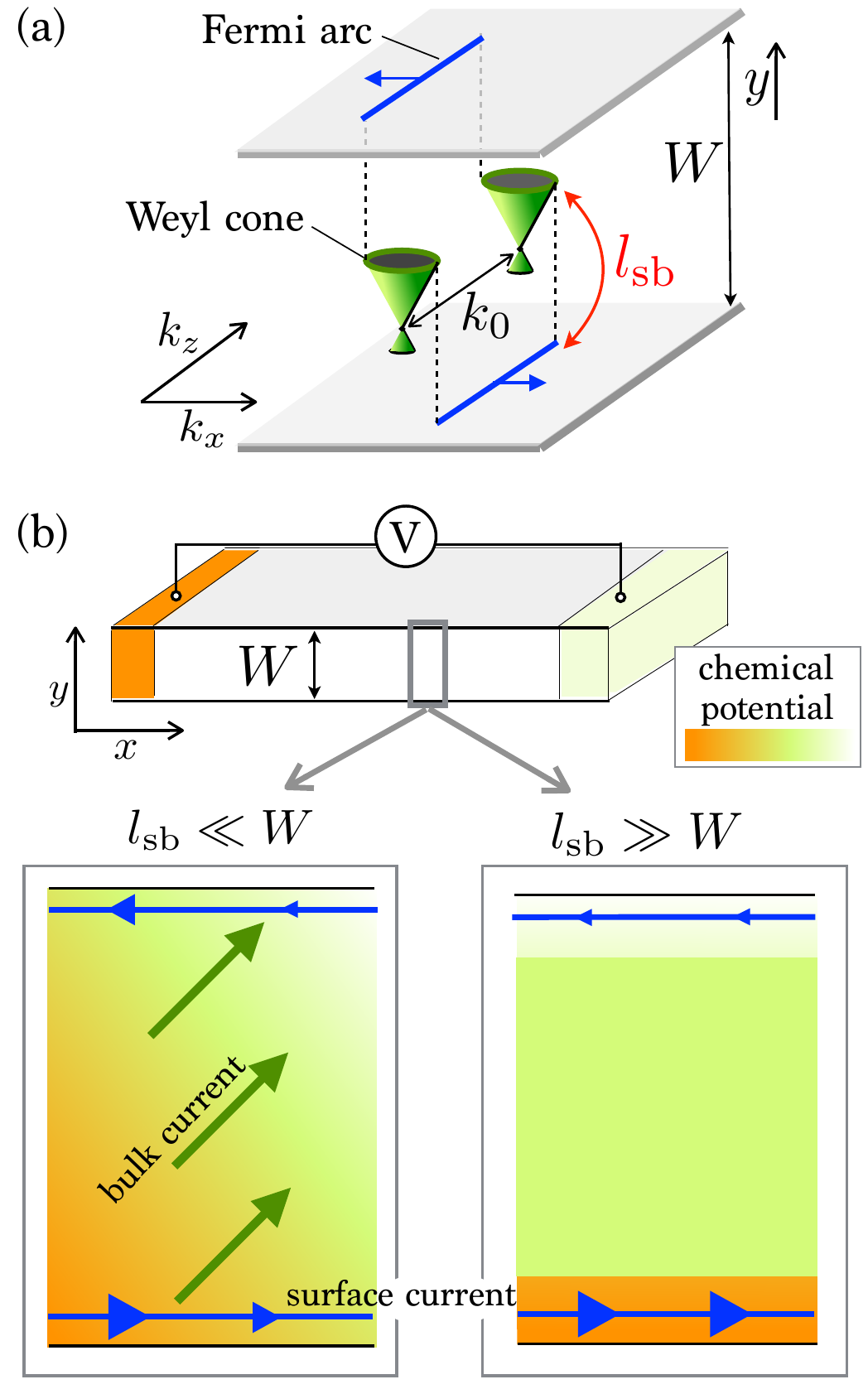}
\caption{(a) Weyl semimetal with two Weyl cones (green) and straight Fermi-arc surface states (blue)
illustrated in mixed momentum/real space.
(b) A Weyl semimetal in slab geometry (finite in $y$ direction) 
with an imposed potential gradient in the $x$ direction. The enlarged view illustrates the 
chemical-potential gradients of surface and bulk states and the resulting current flow in case
of strong surface bulk scattering $l_\tsb\ll W$ (left) and weak surface-bulk scattering $l_\tsb\gg W$ (right).   }
\label{fig1}
\end{figure}

\textit{Finite-size anomalous Hall effect.} --- For a finite system with dimensions
 $-L_x/2 < x < L_x/2$, $-W_y/2 < y < W_y/2$, $- L_z/2 < z < L_z/2$, 
the non-trivial topology of the 
band structure implies the presence of surface states, which reside on ``Fermi arcs'', that
connect the Weyl-fermion Fermi surfaces in momentum space
\cite{Armitage2017, Balents2011}, as illustrated in Fig.\ \ref{fig1}(a). 
We consider a minimal model of two Weyl nodes, for which
the Fermi arcs are straight lines of length $k_0$ 
and the velocities 
$\mathbf{v}_{\ts} = \hbar^{-1} \nabla_{\mathbf{k}_{\ts}} \varepsilon$ are 
$\mathbf{k}$-independent \cite{*[{The assumption that the Fermi arcs are straight lines and that the 
velocities are $\mathbf{k}$-independent are not essential for our conclusions,
see supplemental material}] [{}] dummy2}. 
In particular, the surface-state velocities are $\pm v\hat{\mathbf{x}}$ and
$\pm v\hat{\mathbf{y}}$ at the surfaces at $y=\mp W_y/2$ and
$x=\pm L_x/2$ respectively.  
The density of Fermi-arc states at each of those four surfaces  is 
\begin{equation}
n_\mathrm{s}= \frac{1}{(2\pi)^2}\int_\text{s} d^2 k \delta (\varepsilon-\varepsilon_F) = \frac{k_0}{2\pi vh}.
\label{ns}
\end{equation}
To determine the conductivity of the finite-size system we follow the Landauer approach and 
interpret the applied electric field $\mathbf{E}$ as a chemical-potential difference between the system 
boundaries.
In this picture, the bulk contribution $\mathbf{j}_\tb$ to the current-density misses any anomalous 
transverse terms from a momentum-shifting $\mathbf{E}$-field \cite{Xiao2010} and instead follows
the standard relation
\begin{align}
\mathbf{j}_\tb  =&   e^2 n_\tb D\, \mathbf{E}.
\label{jb1}
\end{align}
Opposite surfaces have counterpropagating surface states, so that the surface contribution $\mathbf{j}_\ts$ to the current density is driven by the transverse potential differences $E^x L_x$ and $E^y W_y$,
\begin{align}
  j_\ts^x  =&   e^2 n_\ts v\, E^y,\;\;\;\;\;\; j_\ts^y  =  -e^2 n_\ts v\, E^x.
\label{js1}
\end{align}
Taken together, Eqs.\ \eqref{ns}--\eqref{js1} reproduce the conductivity tensor \eqref{cond0}. 
We note that the contribution of the anomalous (topological) surface states to the average 
current density of the three-dimensional sample is not antiproportional to the system size, 
unlike that of non-topological surface states.
Owing to spatial separation of countermovers, the non-equilibrium occupation of chiral 
surface states is proportional to $L_x$ or $W_{y}$, which cancels with the same factor,
by which the total current is divided to obtain the current density. 

In the above derivation we assumed local equilibrium between surface and bulk states, 
something that is only justified if the coupling between bulk and surface is sufficiently 
strong \cite{Ye2018, Vinkler-Aviv2018}. To describe a finite coupling strength between
bulk and surface states, we introduce
the relaxation length $l_\tsb$, corresponding to elastic isotropic scattering between surface and bulk states,
neglecting any material-specific dependence of the scattering amplitude on the scattering states \cite{Resta2018}. We consider a slab geometry, for which $L_{x}$ and $L_z$ are taken to infinity, whereas $W_y \equiv W$ is finite, see Fig.\ \ref{fig1}(b). (Here ``infinite'' means $L_{x}$, $L_z \gg l_\tsb $). The current is applied in the $x$ direction.
The (surface) charge densities $c_{\ts\pm}$ of surface states at $y=\pm W/2$ and the (bulk) charge density of the bulk states read, respectively, 
\begin{align}
c_{\ts\pm}=-en_\ts\mu_{\ts\pm},\ \ c_\tb=-e n_\tb\mu_\tb,  \label{ci}
\end{align}
where $\mu_{\ts\pm}$, $\mu_{\tb}$ is the corresponding (local) deviation of the chemical potentials 
from the Fermi energy. We assume that the penetration depth of surface states is much smaller 
than all other relevant length scales.
Introducing surface and bulk current densities $\mathbf{j}_{\ts\pm}=j_{\ts\pm}\hat{\mathbf{x}}$ and 
$\mathbf{j}_\tb$, the set of equations that determine the non-equilibrium steady state reads
\begin{subequations}
\begin{align}
\mathbf{j}_\tb=& -D\boldsymbol{\nabla} c_\mathrm{b}, \label{jb}\\
j_{\ts\pm}=& \, \mp v \, c_{\ts\pm},\label{js}\\
\boldsymbol{\nabla}\cdot\mathbf{j}_\tb =& - \sum_{\pm}
  \partial_x j_{\ts\pm} \delta(y\mp W/2), \label{djb}\\
\partial_x j_{\ts\pm} =& -e n_\ts v \frac{\mu_b(\pm W/2)-\mu_{\ts\pm}}{l_\tsb}.\label{djs}
\end{align}
\label{e1}
\end{subequations}%

\noindent
The first equation describes the diffusion of bulk particles with the diffusion constant $D$. 
The charge current density of chiral surface particles is directly proportional to their 
charge density. The change of the 
current densities is related to scattering between bulk and surface. This is captured by the 
continuity equation  \eqref{djb} and by Eq.\ \eqref{djs}, which relates the rate of change of
$c_{\ts\pm}$ to the chemical potential difference $\mu_b-\mu_{\ts\pm}$ and the scattering
length $l_{\tsb}$. The surface-state penetration depth being much smaller than $W$
excludes direct scattering between surface states of opposite surfaces.

\begin{figure}[t]
\includegraphics[scale=0.7]{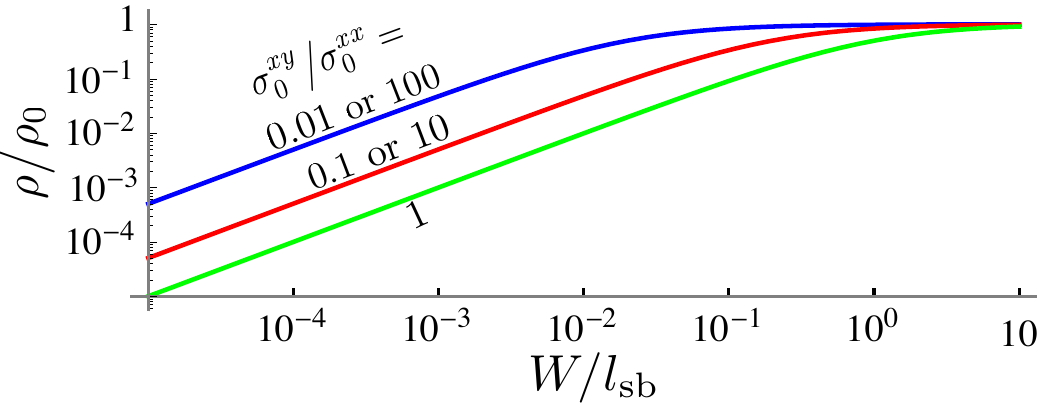}
\caption{Resistivity as a function of the slab width for different values of
$\sigma_0^{xy}/\sigma_0^{xx}$. With decreasing $W/\l_\tsb $ the resistivity goes to zero
 since the current is increasingly conducted via surface states, which dissipation decreases. }
\label{fig2}
\end{figure}

Translation invariance in the $x$-direction allows us to assume a constant gradient of the 
chemical potential, corresponding to an applied electric field $E$ in the $x$ direction. To 
linear order in the gradients, we thus may set
\begin{equation}
\mu_\tb=eEx+\tilde{\mu}_\tb(y)\;\;\;\;\;\; \mu_{\ts\pm}=eEx+\tilde{\mu}_{\ts\pm}. \label{mui}
\end{equation}
Inserting this ansatz into \eqref{e1} gives $j_\tb= \sigma_0^{xx} E$ and
\begin{align}
\tilde{\mu}_\tb(y)=eE_{\perp} y, \ \
\tilde{\mu}_{\ts\pm}=\pm eE_{\perp} \frac{W}{2}\pm eE\, l_\tsb,
\label{musy}
\end{align}
where $E_{\perp} = \sigma^{xy}_0 E/\sigma^{xx}_0$ and where $\sigma_0^{xx} = e^2D n_\tb$ and $\sigma_0^{xy} = e^2v n_\ts$ are the components of the infinite-system conductivity \eqref{cond0}. 
The combined current density $j_\ts = (j_{\ts+} + j_{\ts -})/ W$ coming from surface states is then 
\begin{align}
j_\ts =&  \sigma_0^{xy} \Bigg( \frac{\sigma_0^{xy}}{\sigma_0^{xx}}+\frac{2l_\tsb}{W}\Bigg)E.
\label{js2}
\end{align}
In the limit $W\gg l_\tsb $ we recover the infinite-system conductivity \eqref{cond0} by
inverting the resistivity tensor $\rho^{xx} = \rho^{yy} = E/(j_{\ts} + j_{\rm b})$
and $\rho^{yx} = -\rho^{xy} = E_{\perp}/(j_{\ts} + j_{\rm b})$. For finite
$l_\tsb /W$, the transverse resistivity is not well defined due to the chemical-potential difference 
between surface and bulk states --- it depends on to which subsystem the apparatus measuring the Hall 
voltage couples \cite{Ye2018, Vinkler-Aviv2018}. The longitudinal resistivity remains well defined,
however, see Fig.\ \ref{fig2}. It includes the finite-size term $2 l_\tsb/W$, which grows for weak 
surface-bulk coupling and pushes the current density profile towards the surfaces, as shown in 
Fig.\ \ref{fig1}(c). 
Decreasing the slab width below $ l_\tsb/ |\sigma_0^{xy}/\sigma_0^{xx}
+\sigma_0^{xx}/\sigma_0^{xy}|$ the current flows with less and less dissipation --- more and more
current is carried by the surface states --- and
the resistivity goes to zero. Note that
$\rho/\rho_0$ is an even function of $\sigma_0^{xx}/\sigma_0^{xy}$ and
$\sigma_0^{xy}/\sigma_0^{xx}$, hence the resistivity has a generic minimum at $\sigma_0^{xy}/\sigma_0^{xx}=1$ 
for fixed $W/l_\tsb$. 

\begin{figure}[b]
\includegraphics[scale=0.45]{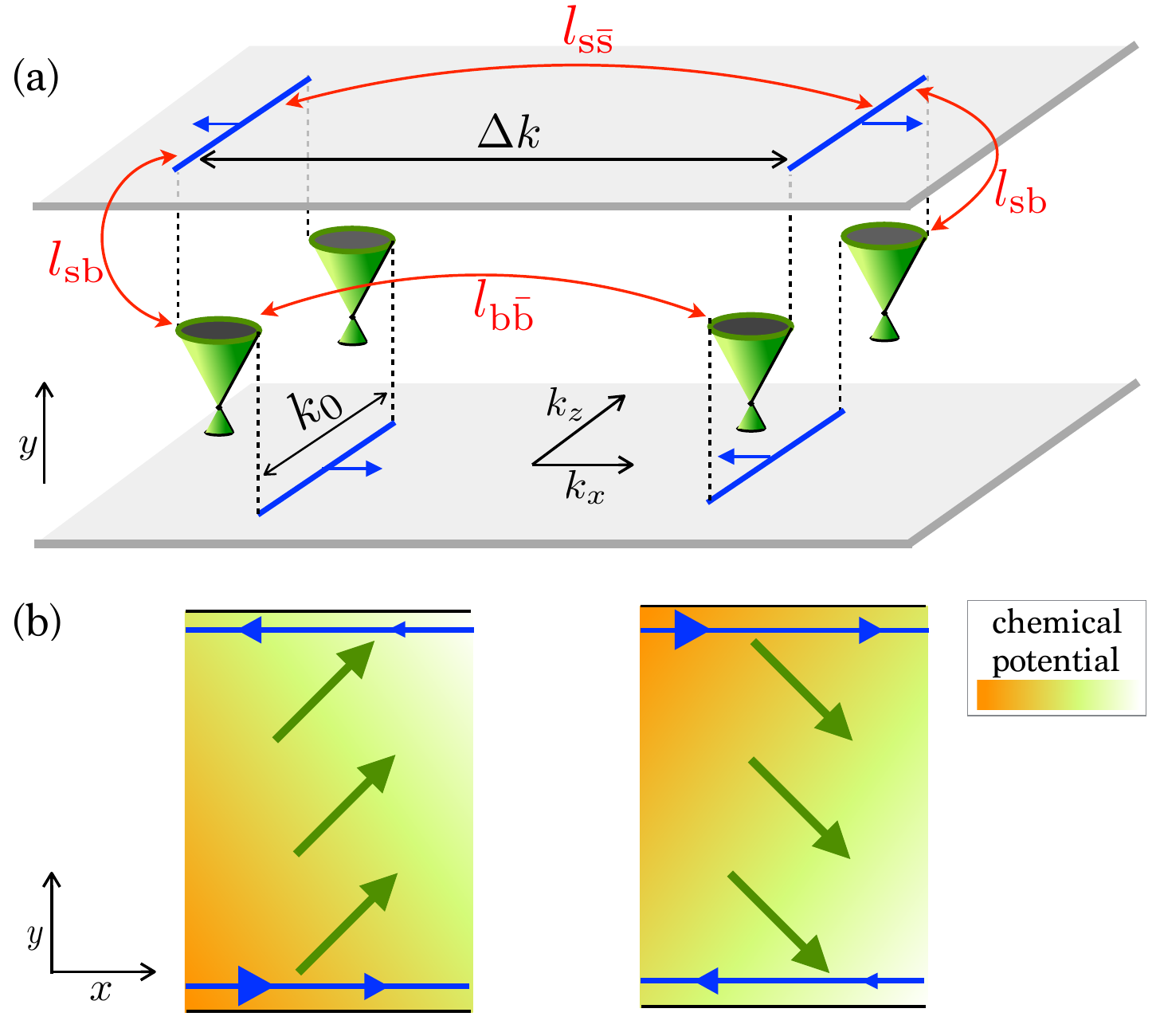}
\caption{(a) TR-symmetric Weyl semimetal build out of two TR copies of the previous model [cf.\ Fig.\ \ref{fig1}(a)]
separated by $\Delta k$ and coupled by scattering (quantified by the relaxation length $l_\tss$ and $l_\tbb$). 
(b) Potential gradients and current flow of the two subsystems for $\l_\tsb\ll W \ll \bar l =\min (l_\tss,\sqrt{2Dl_\tbb/v})$.
Both subsystems exhibit the 
anomalous Hall effect of a TR broken Weyl semimetal [cf.\ Fig.\ \ref{fig1}(c)]. The Hall voltages cancel each other 
but the surface-state currents add up.}
\label{fig3}
\end{figure}

\textit{Unbroken time-reversal symmetry}. --- To describe TR symmetric Weyl semimetals we extend
the two-node minimal model discussed above by adding its time-reversed copy and separating 
the copies in momentum space by $\Delta k$, as illustrated in 
Fig.\ \ref{fig3}(a). 
Using $\bar \ts$ and $\bar \tb$ to denote surface states and bulk states in the time-reversed copy,
we introduce relaxation length $l_\mathrm{b \bar b}$ and $l_\mathrm{s \bar s}$ for 
scattering between bulk and surface states of different TR subsystems. As before, $l_{\tsb}$ is
the relaxation length for scattering between surface and bulk states in the same TR subsystem.
%For simplicity we
%do not consider scattering processes in which surface states from one TR subsystem scatter into
%bulk states of the other TR subsystem and vice versa ({\em i.e.}, $\rm b \leftrightarrow \bar s$ and $\rm \bar b \leftrightarrow s$), 
%although we note that our theory can be
%easily extended to include such processes.
 Scattering processes of the type $\ts \leftrightarrow \rm \bar b$, which connect surface and bulk states of different TR subsystems, are expected to be weaker than processes of the type $\ts  \leftrightarrow \rm \bar s$, because the density of bulk states at the surface is much smaller than the density of surface states, as seen, {\em e.g.}, in ARPES experiments \cite{Xu2015a, Lv2015, Liu2016}. For simplicity, we here neglect these scattering processes completely, although we note that our theory can be easily extended to include them.

 The proper generalization of Eqs.\ \eqref{djb} and 
\eqref{djs} then reads
\begin{subequations}
\begin{align}
\boldsymbol{\nabla}\cdot \mathbf{j}_\tb =&\, \sum_{\pm} e n_\ts v \delta(y\mp W/2)\frac{\mu_b-\mu_{\ts\pm}}{l_\tsb} 
+v\frac{c_{\bar{\tb}}-c_\tb }{l_\mathrm{b \bar b}},\label{djb2}\\
\partial_x j_{\ts\pm} =&\, - e n_\ts v \frac{\mu_b(\pm W/2)-\mu_{\ts\pm}}{l_\tsb}
+ v \frac{c_{\bar{\ts}}-c_\ts}{l_\mathrm{s \bar s}},\label{djs2}
\end{align}%
\label{trseq}
\end{subequations}%

\noindent
plus two equations for $\boldsymbol{\nabla} \cdot \mathbf{j}_{\bar \tb}$ and $\partial_x j_{\bar \ts \pm}$. 
Solving these equations (see supplemental material for details)
gives $j_\tb = 2 \sigma_0^{xx}$ and 
\begin{align}
j_\ts =& \,\frac{2 \sigma_0^{xy} l_\tss^2}{W(l_\tss+2 l_\tsb)}\label{resjs0}
 \\
&\, \mbox{} \times \Bigg[ \frac{\sigma_0^{xy} \ls}{\sigma_0^{xy} \ls
+\sigma_0^{xx} (l_\tss + 2 l_\tsb) \coth (W/\ls)}
+2 \frac{l_\tsb}{l_\tss}\Bigg]\, E, \nonumber
\end{align}
where $\ls =\sqrt{2Dl_\tbb/v}$. In the limit $l_\tsb\ll l_\tss,\, W$ Eq.\ \eqref{resjs0} simplifies to
\begin{align}
j_\ts =& 2\frac{(\sigma_0^{xy})^2}{\sigma_0^{xx}}\,
\bigg(\frac{W}{l_\tss}\,\frac{\sigma_0^{xy}}{\sigma_0^{xx}}
+\frac{W}{\ls} \coth\frac{W}{\ls}\bigg)^{-1}
\, E. \label{resjs1}
\end{align}

\noindent
In the limit $W\ll \bar l \equiv \min(\ls,l_{\tss})$, we obtain twice the current of the previously discussed TR-broken Weyl semimetal
[cf.\ \eqref{js2}], where the presence of a large surface current was associated with the anomalous Hall effect. Here the potential and current pattern of 
the anomalous Hall effect appear in both TR subsystems, as illustrated in Fig.\ \ref{fig3}(b). Added together, the transverse Hall 
fields $\pm j_\ts/(2\sigma_0^{xy})$
cancel, while the longitudinal current contributions of surface states add up.

\begin{figure}[t]
\includegraphics[scale=0.65]{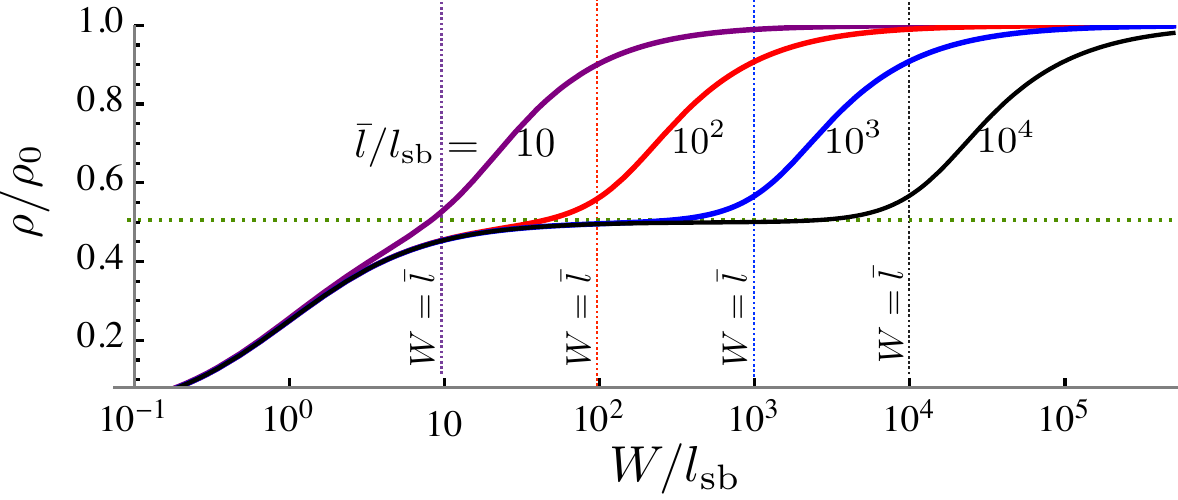}
\caption{Resistivity $\rho$ versus slab width $W$ for different ratios $l^\ast / l_\tsb$ at $\sigma_0^{xy}/\sigma_0^{xx}=1$.
The other parameters were chosen as $l_\tss=l_\tbb$ and $2D/v = l_\tsb$ so that $\bar l = \min(\ls,l_{\tss}) = \ls$. If $l_{\rm sb} < \bar l$, the anomalous contribution of surface states leads to a drop from $\rho/\rho_0=1$ for $\bar l \ll W$ to a plateau at $\rho/\rho_0=1/[1+(\sigma_0^{xy}/\sigma_0^{xx})^2]$ (green dotted line) for  $l_\tsb \ll W\ll \bar l$.}
\label{fig4}
\end{figure}

The total resistivity of the slab, including the bulk contribution 
 $j_\tb=2\sigma_0^{xx}E$, shows characteristic signatures of this ``twin anomalous 
Hall regime''. As shown in Fig.\ \ref{fig4}, in the limit $l_\tsb \ll \bar l $ of weak scattering between the two TR subsystems the resistivity $\rho$ drops in two steps upon decreasing the system width $W$: First a drop from $\rho_0 = 1/2 \sigma^{xx}_0$ to $\rho_0/[1+(\sigma_0^{xy}/\sigma_0^{xx})^2]$ at $W \sim \bar l$, followed by a drop to zero at $W \lesssim l_\tsb$. The first drop is due to the contribution of surface states, which sets in when the two subsystems effectively decouple and develop oppositely directed Hall voltages. The second drop then occurs for the same reasons as for the TR-broken case: The current is pushed towards the surfaces, as these become perfectly conducting channels. If the condition $l_\tsb \ll \bar l$ is not met, the conductivity drops to zero in a single step at $W \sim \min(l_{\tsb},l_{\tss})$.

\textit{Discussion}. --- We have shown that Fermi-arc surface states in TR-invariant 
topological metals can contribute to transport if the system size is much larger than the 
lattice constant or the  bulk mean free path, as long as the transverse
system size $W$ is smaller than $\bar l \equiv \min(l_\tss,l^\ast)$, where $l_\tss$ and $\ls = \sqrt{2Dl_\tbb/v}$ are the characteristic length scales describing the coupling between TR-conjugated subsystems.
In this regime, the topological metal is effectively the sum of 
two time-reversed subsystems, which each exhibit an anomalous Hall effect. The longitudinal 
conductivity increases from the sum $\sigma_0 + \sigma_0^T = 2 \sigma_0^{xx}$ of subsystem 
conductivities in the limit $W \gg \bar l$ to twice the inverse longitudinal resistivity $2 (\rho_0^{xx})^{-1} = 2 \sigma_0^{xx} + 2 (\sigma_0^{xy})^2/\sigma_0^{xx}$ if $W \ll \bar l$, where the extra term is the contribution of the Fermi-arc surface states. That one has to 
average the resistivities instead of conductivities can also be understood directly from 
the fact that for each decoupled subsystem the transverse current must vanish, while the 
transverse potential gradient is set by the applied electric field.
Remarkably, the surface contribution to the conductivity does not scale inversely with the width $W$ as long as $W \ll \bar l$ and is \emph{inversely proportional} to $\sigma_0^{xx}$, making it larger for stronger scattering rates. This is in stark contrast to the opposite limit $W \gg \bar l$, where  the surface contribution to the conductivity is proportional to $l_\tsb/W$ (assuming $l_\tsb\ll l_\tss$), thus inversely proportional to both the width and surface-bulk scattering amplitude \cite{Gorbar2016a}.

To estimate the characteristic length $\bar l$ for existing Weyl semimetals
from the TaAs family, we compare it to the bulk 
scattering length $2D/v \sim 1\, \mu m$ given by the bulk transport lifetime and Fermi velocity 
and estimated from resistivity measurements \cite{Zhang2017d} combined with ab-initio calculations \cite{Lee2015}. 
Assuming that scattering is dominated by a Coulomb-disorder potential \cite{Burkov2011a,
Ominato2015}, 
%or acoustic phonons \cite{Sarma2015}, 
the scattering amplitude is suppressed 
with increasing momentum difference $q$ of scattering states $\propto 1/q^2$. 
Consequently, $2D/v$ is dominated by scattering processes  
within Weyl cones --- on momentum-space distances $ k_F\sim 0.01/\mathrm{\AA}$ \cite{Xu2015a, Arnold2016}.
The scattering lengths $l_\tbb$ and $l_\tss$ are instead governed by scattering on distances $\Delta k \sim  1 / \mathrm{\AA}$,
which gives the estimate $\bar l\sim  100\,  \mu m$.  Additionally $\bar l$ can be increased 
if scattering  between time-reversed states involves a spin flip \cite{Xu2016, Inoue2016}
or is dominated by long-ranged Gaussian impurities \cite{Ominato2015}.
 
An experimental signature of the anomalous contribution of Fermi arcs
is the decreasing resistivity with decreasing system size.
The magnitude of this effect grows with increasing subsystem Hall angle $\sigma_0^{xy} / \sigma_0^{xx}$. While the estimation of this quantity for specific materials goes beyond the scope of this work, we note that a large value of $\sigma_0^{xy} / \sigma_0^{xx}$ is likely to occur, given the presence of a large Hall angle in TR-broken Weyl metals \cite{Suzuki2016, Liu2017c}. We also note that the Hall angle may be increased by an enhanced long-ranged disorder, which would decrease $\sigma_0^{xx}$ without 
restricting the width bound $W \lesssim \bar l $.

The predicted decrease of the resistivity with decreasing sample thickness ($\sim 100\,\mu m$)
is in qualitative agreement with recent
 measurements on NbAs nanobelts \cite{*[{}] [{. Note, in particular, section VI of the Supplemental Material.}] Zhang2019}, remarkably contrasting with 
measurements on the Dirac semimetal Cd$_3$As$_2$ \cite{Liang2015, Zhang2017b, Schumann2018}, 
where such a decrease has not been observed.  Note that in Dirac semimetals $\Delta k \to 0$ 
strongly suppressing the characteristic width at which the twin anomalous Hall effect can set in.
%These observations are consistent with the
%theory presented here, taking into account the large and vanishing
%separation of time-reversed states in NbAs and Cd$_3$As$_2$ respectively, 
%enabling the twin anomalous Hall effect in the former but not in the latter.

Other striking signatures and potential applications of the discussed phenomenon lie 
in transport devices that access the conductivity locally.
The peculiar width independence of
 the surface contribution to the \emph{average} conductivity
implies a highly inhomogeneous \emph{local} conductivity --- enhanced by 
$(W/\xi)\,2 (\sigma_0^{xy})^2 / \sigma_0^{xx} $ at the surface, where
$\xi$ is the penetration depth of the surface states. This factor can be large 
even at small Hall angles, since $W$ is 
only bound by $\bar l$, while $\xi$ is typically on the order of the lattice constant. 

An example of how this surface current can be detected, is the Edelstein effect \cite{Edelstein1990}, which 
has been predicted to be large in TaAs \cite{Johansson2018}, due to the strong spin polarization of Fermi arcs
in these materials \cite{Xu2016, Inoue2016}. The current-induced magnetization will inherits the width dependence
of the local Fermi-arc current density, predicted by our work. In particular, this will lead to a  strong enhancement of the Edelstein effect in thin films with $W\lesssim \bar l$.

In this context, it is interesting to note that a wedge geometry, in which
the width $W$ varies in the direction of current flow between the regimes $W\ll\bar l$ 
and $W\gg \bar l$ converts a surface current into uniform bulk current and vice versa, 
dependent on the current direction. Such a device could be an example how the peculiar size dependence
of topological metals can be used for nanotransport circuit design.

\textit{Acknowledgments}. We would like to thank P. Silvestrov, A. Johansson,
C. Timm, and F. Xiu for valuable discussions. This
research was supported by project A02 of the CRC-TR
183 “entangled states of matter” and the Grant
No. 18688556 of the Deutsche Forschungsgemeinschaft
(DFG, German Research Foundation).

\onecolumngrid

\bibliography{library}

%\clearpage

\renewcommand{\theequation}{S\arabic{equation}}
\renewcommand{\thefigure}{S\arabic{figure}}

\setcounter{equation}{0}
\setcounter{figure}{0}

\section*{Supplemental Material}

\section{Current contribution of curved Fermi arcs}

Our analysis can be easily generalized to curved Fermi arcs instead of the straight Fermi arcs considered for simplicity in the main text. In the following we show that the expression for the current density used in the main text given by combination of Eqs.\ \eqref{ns} and \eqref{js1},
\begin{equation}
j_{\ts}^x = \frac{e^2k_0}{2\pi\, h}\,E_y,\;\;\;\;\;\;\;
j_{\ts}^y = -\frac{e^2k_0}{2\pi\, h}\,E_x,
\end{equation}
continues to hold in case of curved Fermi arcs, $k_0$ being the length of a straight line connecting the end points of the arc.   
In case of curved Fermi arcs, the current contribution of Fermi arcs at two opposite surfaces  is given by 
\begin{equation}
\mathbf{j}_{\ts\pm}=-e\,\mu_{\ts\pm}\,\frac{1}{(2\pi)^2}\int d^2k\,\delta (\varepsilon-\varepsilon_F)  \mathbf{v}_{\ts\pm},
\end{equation}
where $\mu_{\ts\pm}$ is the deviation of the chemical potential from the Fermi level.
Executing the integration over energy, the integral reduces to the 1D integration along the Fermi arc, 
\begin{equation}
\hbar \int d^2k\,\delta (\varepsilon-\varepsilon_F)  \mathbf{v}_{\ts\pm} 
= \int dk\, \frac{\mathbf{v}_{\ts\pm}}{|\mathbf{v}_{\ts\pm}|}. \label{int1}
\end{equation}
The integral is solved by closing the integration contour with a straight line of length $k_0$ 
that connects the two end points of the Fermi arc. Let $\mathbf{n}_\pm$ be a unit vector normal to this line, 
such that the angle between 
 $\mathbf{v}_{\ts\pm}\big/|\mathbf{v}_{\ts\pm}|$ and the integration contour is kept constant 
 if we define $\mathbf{v}_{\ts\pm}\big/|\mathbf{v}_{\ts\pm}|=\mathbf{n}_\pm$ on this line. 
We then obtain
\begin{equation}
\int dk\, \frac{\mathbf{v}_{\ts\pm}}{|\mathbf{v}_{\ts\pm}|} 
= \underbrace{\oint dk\, \frac{\mathbf{v}_{\ts\pm}}{|\mathbf{v}_{\ts\pm}|}}_{=0} - k_0\,\mathbf{n}_\pm,
\end{equation}
leading to the result 
\begin{equation}
\mathbf{j}_{\ts\pm}=\frac{e\,k_0}{2\pi\, h}\,\mu_{\ts\pm}\,\mathbf{n}_\pm.
\end{equation}
Applying to the current contribution of the surfaces $y=\pm W_y/2$, where $\mathbf{n}_\pm = \pm \hat{\mathbf{x}}$,
and the surfaces $x=\pm L_x/2$, where $\mathbf{n}_\pm = \mp \hat{\mathbf{y}}$,
we obtain, respectively,
\begin{equation}
j_{\ts}^x = \frac{j^x_{\ts+}+j^x_{\ts-} }{W_y}= \frac{e^2k_0}{2\pi\, h}\,E_y,\;\;\;\;\;\;\;
j_{\ts}^y = \frac{j^y_{\ts+}+j^y_{\ts-} }{L_x}= -\frac{e^2k_0}{2\pi\, h}\,E_x.
\end{equation}

Note that, while the final expression for the current density of surface states remains valid for curved Fermi arcs, the expression for the density of states $n_s$ given in \eqref{ns} is no valid in case of curved Fermi arcs. The expressions for the current density used in the main text, however, remain valid in case of curved Fermi arcs if the product $n_s v$ is replaced by $k_0/2\pi h$.

\section{Unbroken time-reversal symmetry}

In the following we discuss in detail the solution for the TR-preserved case.
From Eqs.\ \eqref{jb}, \eqref{js}, \eqref{djb2}, \eqref{djs2}, \eqref{ci}, and \eqref{mui} we obtain the set of equations
\begin{subequations}
\begin{align}
j_\tb^y=& e n_\tb D\, \partial_y\mu_\tb(y),\label{jbdmu}\\
\partial_y j_\tb^y =& e\frac{n_\ts v}{l_\tsb}\sum_\pm\big[\mu_\tb(y)-\mu_{\ts\pm}\big]\delta\big(y\mp\tfrac{W}{2}\big)+2e\frac{n_\tb v}{l_\tbb}\mu_\tb (y),\\
\pm eE =& - \frac{\mu_\tb(\pm W/2)-\mu_{\ts\pm}}{l_\tsb}+2\frac{\mu_{\ts\pm}}{l_\tss}.
\end{align}
\label{deq3}
\end{subequations}
Eliminating $j_\tb^y$ in \eqref{deq3} and using the symmetry relation
$\mu_\tb\big(\pm\tfrac{W}{2}\big)=\pm\mu_\tb\big(\tfrac{W}{2}\big)$  we obtain
\begin{subequations} \begin{align}
\partial_y^2 \mu_\tb =& \frac{n_\ts v}{n_\tb D}\sum_\pm
\frac{\mu_\tb(y)-\mu_{\ts\pm}}{ l_\tsb }\delta\big(y\mp\tfrac{W}{2}\big)
+\frac{2v}{D\,l_\tbb}\mu_\tb(y),\label{d2mu} \\
\mu_{\ts\pm} =& \pm\frac{ eE\,l_\tsb+\mu_\tb\big(\tfrac{W}{2}\big)}{1+2l_\tsb/l_\tss}. \label{d2mus} 
\end{align} \label{deq5} \end{subequations}

Integration of Eq.\ \eqref{d2mu} with the boundary condition $\mu_\tb'(-W/2)=0$, which according 
to \eqref{jbdmu} corresponds to $j_\tb^y(-W/2)=0$,  and inserting Eq.\ \eqref{d2mus} gives 
\begin{align}
\mathrm{lim}_{\xi\to 0}\,\mu_\tb'(-W/2+\xi)=& \mathrm{lim}_{\xi\to 0}\,\int_{-W/2}^{-W/2+\xi}dy \,\partial_y^2 \mu_\tb 
 =  \frac{n_\ts v}{n_\tb D}\frac{\mu_\tb(-W/2)-\mu_{s-}}{l_\tsb} \nonumber \\
=& \frac{n_\ts v}{n_\tb D}\bigg[eE+\frac{2\mu_\tb(-W/2)}{l_\tss+2 l_\tsb}\bigg].\label{BC1}
\end{align}
The solution of \eqref{d2mu} for $-W/2<y<W/2$ with $\mu_\tb(-y)=-\mu_\tb( y)$ reads
\begin{equation}
\mu_\tb( y) = A\,\Big(e^{-2(y+W/2)/l^\ast}-e^{2(y-W/2)/l^\ast}\Big),\;\;\;\; l^\ast=\sqrt{2Dl_\tbb/v}.
\end{equation}
The integration constant $A$ is determined by the boundary condition \eqref{BC1}, giving the solution
\begin{equation}
\mu_\tb( y) = eE\,\frac{W}{2}\,\frac{1}{1+2l_\tsb/l_\tss}\,
\frac{e^{2(y-W/2)/l^\ast}-e^{-2(y+W/2)/l^\ast}}{\frac{W}{l_\tss+2l_\tsb}
\big(1-e^{-2 W/l^\ast}\big)+\frac{ W}{l^\ast}\big(1+e^{-2 W/l^\ast}\big)\frac{n_\tb D}{n_\ts v}}.
\end{equation}
Using Eq.\ \eqref{d2mus} we obtain 
\begin{align}
\frac{\mu_{s+}-\mu_{s-}}{W} =& eE\,\frac{1}{1+2 l_\tsb/l_\tss}
\Bigg[ \frac{1}{1+2l_\tsb/l_\tss}\,
\frac{1}{\frac{W}{l_\tss+2l_\tsb}
+\frac{n_\tb D}{n_\ts v}\, \frac{W}{l^\ast} \coth\tfrac{W}{l^\ast}}
+\frac{2 l_\tsb}{W}\Bigg].
\end{align}
The current-density contribution of Fermi arcs is given by $j_\ts=2e vn_\ts (\mu_{\ts +}-\mu_{\ts -})/W$, 
where the factor $2$ accounts for the equivalent contribution of the TR subsystem,
\begin{align}
j_\ts =& 2 \sigma_0^{xy}\,\frac{1}{1+2 l_\tsb/l_\tss}
\Bigg[ \frac{1}{1+2l_\tsb/l_\tss}\,
\frac{1}{\frac{W}{l_\tss+2l_\tsb}
+\frac{\sigma_0^{xx}}{\sigma_0^{xy}}\, \frac{ W}{l^\ast} \coth\tfrac{ W}{l^\ast}}
+\frac{2 l_\tsb}{W}\Bigg]\, E.
\end{align}
This is Eq.\ (\ref{resjs0}) of the main text.
In the limit $l_\tsb\ll l_\tss,\; W$ this simplifies to
\begin{align}
j_\ts =& 2\frac{(\sigma_0^{xy})^2}{\sigma_0^{xx}}\,
\bigg(\frac{W}{l_\tss}\,\frac{\sigma_0^{xy}}{\sigma_0^{xx}}
+\frac{ W}{l^\ast} \coth\frac{ W}{l^\ast}\bigg)^{-1}
\, E.
\end{align}
The $W$-dependent term goes to $1$ when $W\to 0$. Hence the twin anomalous Hall regime, in which 
the current-density contribution of surface states is  $W$-independent,
$j_\ts = 2(\sigma_0^{xy})^2 / \sigma_0^{xx}\, E$, is characterized by 
\begin{equation}
\frac{W}{l_\tss}\,\frac{\sigma_0^{xy}}{\sigma_0^{xx}}
+\frac{ W}{l^\ast} \coth\frac{ W}{l^\ast} \lesssim 1, \label{cond}
\end{equation}
which, assuming $\sigma_0^{xy}/\sigma_0^{xx}\sim 1$, 
is equivalent to $W \lesssim \min(l_\tss,l^\ast)$.

\end{document}